\documentclass[letterpaper, 10 pt, conference]{ieeeconf}

\IEEEoverridecommandlockouts

\usepackage{graphicx}
\usepackage{textgreek}
\usepackage{amsmath}
\usepackage[font=footnotesize]{caption}
\usepackage{subcaption}
\usepackage{booktabs}
\usepackage{multirow}
\usepackage{array}
\usepackage{cite}
\usepackage{algorithmic}

\usepackage[ruled,vlined,]{algorithm2e}
\usepackage{makecell}
\usepackage{diagbox}
\usepackage{multirow}

\title{\LARGE \bf
SwarmPlay:  Interactive  Tic-tac-toe  Board  Game  with Swarm  of  Nano-UAVs driven by Reinforcement Learning
}

\author{Ekaterina Karmanova, Valerii Serpiva, Stepan Perminov, Aleksey Fedoseev, and Dzmitry Tsetserukou
\thanks{The authors are with the Intelligent Space Robotics Laboratory, Space CREI, Skolkovo Institute of Science and Technology, Moscow, Russian Federation.
 {\tt \{ekaterina.karmanova, valerii.serpiva, stepan.perminov, aleksey.fedoseev, d.tsetserukou\}@skoltech.ru}}
}

\begin{document}

\maketitle
\thispagestyle{empty}
\pagestyle{empty}


\begin{abstract}

Reinforcement learning (RL) methods have been actively applied in the field of robotics, allowing the system itself to find a solution for a task otherwise requiring a complex decision-making algorithm. In this paper, we present a novel RL-based Tic-tac-toe scenario, i.e. SwarmPlay, where each playing component is presented by an individual drone that has its own mobility and swarm intelligence to win against a human player. Thus, the combination of challenging swarm strategy and human-drone collaboration aims to make the games with machines tangible and interactive. Although some research on AI for board games already exists, e.g., chess, the SwarmPlay technology has the potential to offer much more engagement and interaction with the user as it proposes a multi-agent swarm instead of a single interactive robot. We explore user's evaluation of RL-based swarm behavior in comparison with the game theory-based behavior.
The preliminary user study revealed that participants were highly engaged in the game with drones (70\% put a maximum score on the Likert scale) and found it less artificial compared to the regular computer-based systems (80\%). The affection of the user's game perception from its outcome was analyzed and put under discussion. User study revealed that SwarmPlay has the potential to be implemented in a wider range of games, significantly improving human-drone interactivity. 

\end{abstract}

\section{Introduction}

\begin{figure}[!h]
 \includegraphics[width=1.0\linewidth]{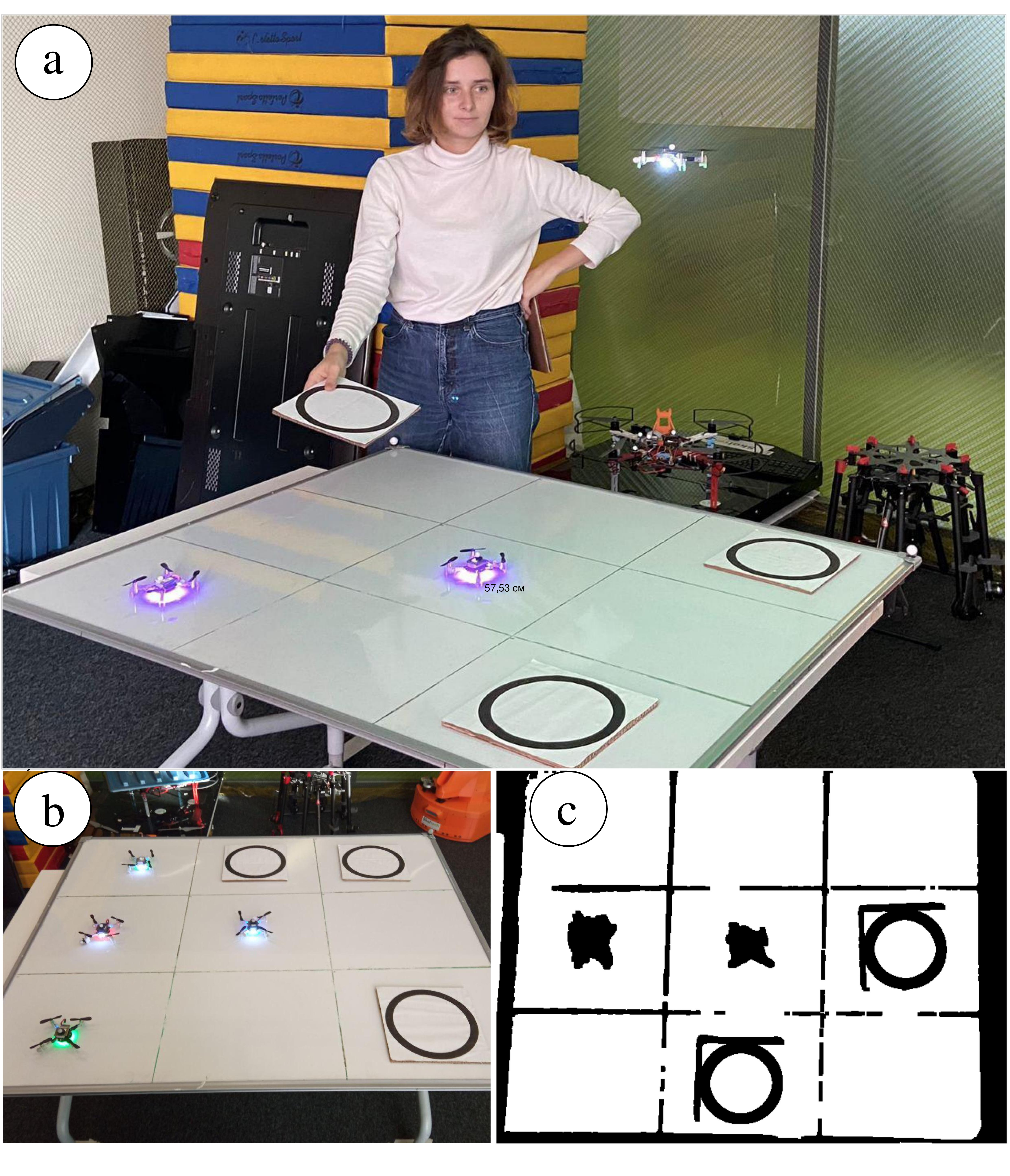}
 \caption{Tic-tac-toe game with swarm of drones: (a) A human plays Tic-tac-toe board game against the SwarmPlay. (b) A closed view on the board when drones won the match. (c) Game board image processed for the Computer Vision (CV) analysis.}
 \label{fig:main}
\vspace{-1.5em}
\end{figure}

One of the potential domains for the human-robot interaction research is physical board games with an adjustable structure level. Perceiving the game components and the board, understanding human movements, reasoning about the state, and manipulating the game components to win against human players are integral steps in robot-centric board games \cite {Li_Differential, matuszek2011gambit}. For the human player, on the other side, the interaction with a robot provides a fresh perspective on the well-known competitive games, e.g., robotic rock-paper-scissors with RASA presented by Ahmadi et al. \cite{Ahmadi_Rock_Paper}.

Today much work is aimed to improve the AI in such games. On the one hand, the researchers achieve a high level of AI by improving the estimation of human behavior, e.g., tennis player's movement prediction proposed by Wu et al. \cite{Wu_FuturePong}. 
Another approach is to develop systems with a sufficient game strategy. RL is one of the most frequently used machine learning approaches in robotics, among other AI algorithms, due to its relative simplicity in comparison with Deep Learning algorithms and a variety of other Supervised learning algorithms, which require much more sophisticated and detailed instructions on problem-solving. Meanwhile, RL algorithms are the ones that react to an environment and finds a solution on their own, by trial-and-error, and do not require any pre-defined dataset to be trained. Therefore, RL is widely implemented in in-game scenarios, e.g., online-game strategy Dota 2 \cite{silva2017moba}, gaming decision system developed for Go by Silver et al. \cite{Silver2016MasteringTG}, and curling robot with adaptive deep reinforcement learning framework proposed by Won et al. \cite{Woneabb9764}. 
However, the system architecture in robot-centric applications has been relatively little investigated and is narrowed to the single robotic manipulators and mobile robots \cite{nugroho2014design, Kyohei_PP_Robot, Becker_Chess}. The research on multi-robot games, though, is mostly focused on coordination between robotic agents, such as soccer game strategies suggested by Reis et al. \cite{Reis_multiSoccer} and Liu et al. \cite{Liu_Soccer} that exclude humans from the gaming stage.
There are several works introducing novel human-drone interaction (HDI) approaches in various scenarios. For example, the Flyables system by Knierim et al. \cite{knierim2018flyables} presents nano-quadrotors as levitating tangibles in 3D space which can be controlled by the user. A similar approach is presented in BitDrones by Gomes et al. \cite{gomes2016bitdrones}, where nano-quadrotors are used as self-levitating tangible building blocks, forming an interactive 3D display with a touchscreen array.
GridDrones is another multi-agent system where the user directly interacts with a volumetric mid-air grid of 15 cube-shaped nano-quadrotors \cite{braley2018griddrones}. A novel system SwarmCloak introduced by Tsykunov et al. \cite{SwarmCloak} proposes the landing of a swarm of four flying nano-quadrotors on the light-sensitive pads with vibrotactile feedback attached to the user's arms. Another approach for drone interaction was suggested in DroneLight developed by Ibrahimov et al. \cite{DroneLight}. The proposed drone control system is based on the gesture recognition, followed by the drone light-painting of predefined patterns collocated with human gestures.
Nitta et al. proposed a novel gaming approach HoverBall with a flying ball based on quadrotor technology, allowing to change the ball's physical dynamics and behavior based on the context of the sports game \cite{HoverBall}. 


 To upgrade the level of engagement and interactivity of traditional games, we suggest a novel game paradigm where each game piece has its own intelligence and mobility, and behaves jointly with other agents to win against the opponent (see Fig. \ref{fig:main}). The proposed SwarmPlay technology provides human-swarm interaction (HSI) driven by RL in board games. To our knowledge, our prototype is the first approach towards using a multi-UAV system in physical games that involves human presence. In this research, we focus on the system architecture and its validation by user study, followed by a discussion about future work and potential SwarmPlay game applications. 

\section{System Overview} 
\subsection{System Architecture} 

\begin{figure} [h]
\centering
\includegraphics[width=1\linewidth]{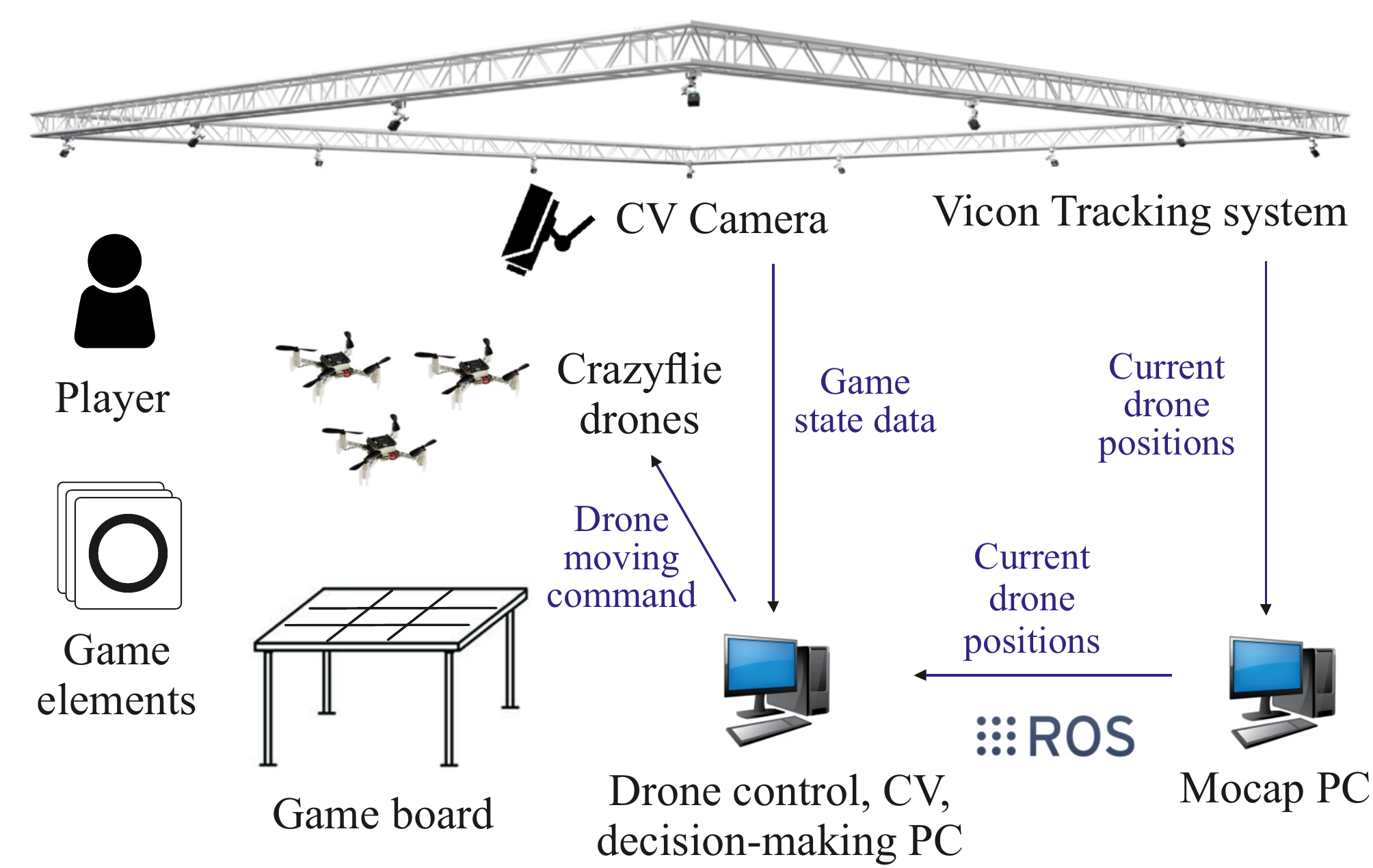}
 \caption{Swarm play game layout. Human player performs their turn with cardboard game elements, then the system estimates the current state of the game with the developed CV system and calculates next command for the drones.}
 \label{fig:Overwiev}
\end{figure}

The developed SwarmPlay system consists of Crazyflie drones, Vicon Tracking system with 12 IR cameras for drone localization, a CV camera for the game state evaluation, a drone landing table with a game board, and PCs with Mocap framework, drone-control framework, CV system, and decision-making system (Fig. \ref{fig:Overwiev}). The game board is divided into 9 cells according to Tic-tac-toe rules. 
According to the specified algorithm, the drones play Crosses (Xs), landing on the game board's cells.
At the same time, a human plays Noughts (Ox), placing cards with printed circles on the game board.

To obtain pictures of the game board providing awareness of a current status of the game, we use a camera Logitech HD Pro Webcam C920 of @30FPS mounted on the ceiling of the room. The game board is placed right under the camera. The pictures are sent to the CV system to determine the human's turn. After that, data on the human's turn as a cell number is sent to the decision-making system to define a cell where the drones should make their next turn. CV and decision-making processing is performed on Intel® Core™ i7-9750HF CPU @ 2.60GHz × 12 threads of execution. The most recent cell data is sent to the drone-control framework. The framework obtains both the target cell, where a drone should be sent, and data from the motion capture system about current drone positions. To obtain the high-quality tracking of the drones, we applied Vicon motion capture system with 12 cameras (Vantage V5) covering a 27 m${^3}$ space. Drones are sent to the target cells with PID control parameters, i.e., the target position, speed, and acceleration. The Robot Operating System (ROS) Kinetic framework is applied to run the developed software and ROS stack \cite{HoenigMixedReality2015} for Crazyflie 2.0. The position and altitude are updated at 100 Hz for all drones. 


To detect to which cell the user puts circles, we developed a corresponding CV system. As its input, we use a picture taken by an RGB camera mounted on the ceiling of the room where the gaming board is located. At each step, the CV system takes a picture and converts it to the grayscale. Then, thresholding and erosion with a kernel 5x5 are applied.
After that, the picture is cropped and divided into 9 small images, one per game cell. For each small image, a contour search is performed. When users make their turn, they put a circle on a cell, which is then detected as a contour by the CV system and filled with black pixels.
At the end of each step, the CV system computes the density of black pixels per game cell. In this case, big colored circles show a great density value. Thus, using some threshold, makes it possible to separate game elements, drones, and empty areas from each other.
After detecting a new circle on the playing field, the CV system sends a corresponding game cell number, as the latest human turn, to a decision-making system to solve how exactly drones should behave in the situation.

\section{Game Strategy}

\subsection{Implementation}


The Tic-tac-toe game is played on a three-by-three grid. Each player takes a turn to place a symbol on an open square. The drones play as an “X" player, and the user is playing as an “O" player. The game is over if one of the players has three identical elements in a row: horizontally, vertically, or diagonally. The game can end with a draw result if there is no possibility to achieve any winning combination. The board is represented by two-dimensional 3x3 matrix 
, where each cell was enumerated as 1, 2, 3, ... 9. Each element of the matrix equals one of the following values:
0 : Unoccupied Cell; +1: Drone Symbol “X"; -1: Player symbol “O".




\subsection{Reinforcement Learning Algorithm}

Most game theory algorithms applied in human-robot interaction scenarios, e.g., Minmax algorithm \cite{Karamchandani}, presume an opponent (human) and an agent (robot) to be ideal. Thus, during the game, each step of the opponent's strategy aims to maximize their reward and minimize the agent's reward, while the agent actions (next game steps) aim to avoid the loss outcome. However, in the Model-free RL algorithms, the agent does not operate with any model of an opponent, nevertheless achieving high results.

\begin{figure}[h!]
\centering
\includegraphics[width = 1\linewidth]{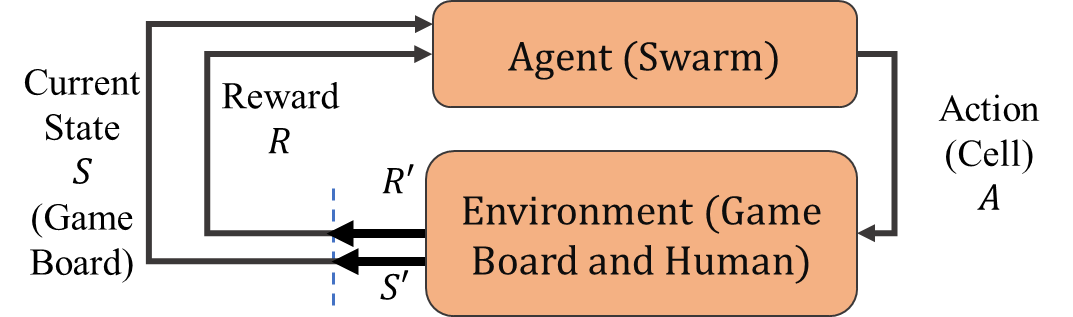}
\caption{RL algorithm: SwarmPlay structure.
}
\label{fig:RL_my}
\end{figure}

In our case, an agent is the swarm of drones, which interacts with an environment being the game board and the opponent, i.e., a human player. Depending on a certain state, i.e., a certain layout of the game board, the swarm takes a certain action by selecting a cell for the next turn. 
The general architecture of the RL algorithm is presented in Fig. \ref{fig:RL_my}, where $S_t$, $S_{t+1}$ is the state at the time $t$ and the next time step $t+1$;
$R_{t}$, $R_{t+1}$ is the reward received by a swarm for achieving the states $S_{t}$, $S_{t+1}$; $A_t$ is the action taken at time $t$ from state $S_t$. The key goal of RL is to define the best sequence of decisions which may allow the swarm to solve a problem while maximizing a long-term reward. 


Temporal Difference (TD) learning is one of the most frequently encountered model-free learning algorithms in RL as it enables the agent to learn through every single action it takes. TD updates the agent's knowledge on every time-step (action) rather than on every episode (reaching the goal or end state). 
In the basic scenario, the state-value (SV) function V(s) is initially evaluated. The agent always takes an action that leads to the state with the highest value. 
Another approach is to evaluate the action-state Q function, i.e., the value of an action in a particular state under a certain policy. 
Depending on how Q-function is updated after each action, the TD methods are subdivided into several approaches to policy learning: 
\begin{itemize}
  \item Q-learning (QL): an off-policy method, where the RL agent learns about policy from another policy.
  \item SARSA: an on-policy method, where the RL agent learns from both current and propagated by one step state and action.
\end{itemize}

\begin{figure}[htbp]
  \centering
  {\includegraphics[width=1\linewidth]{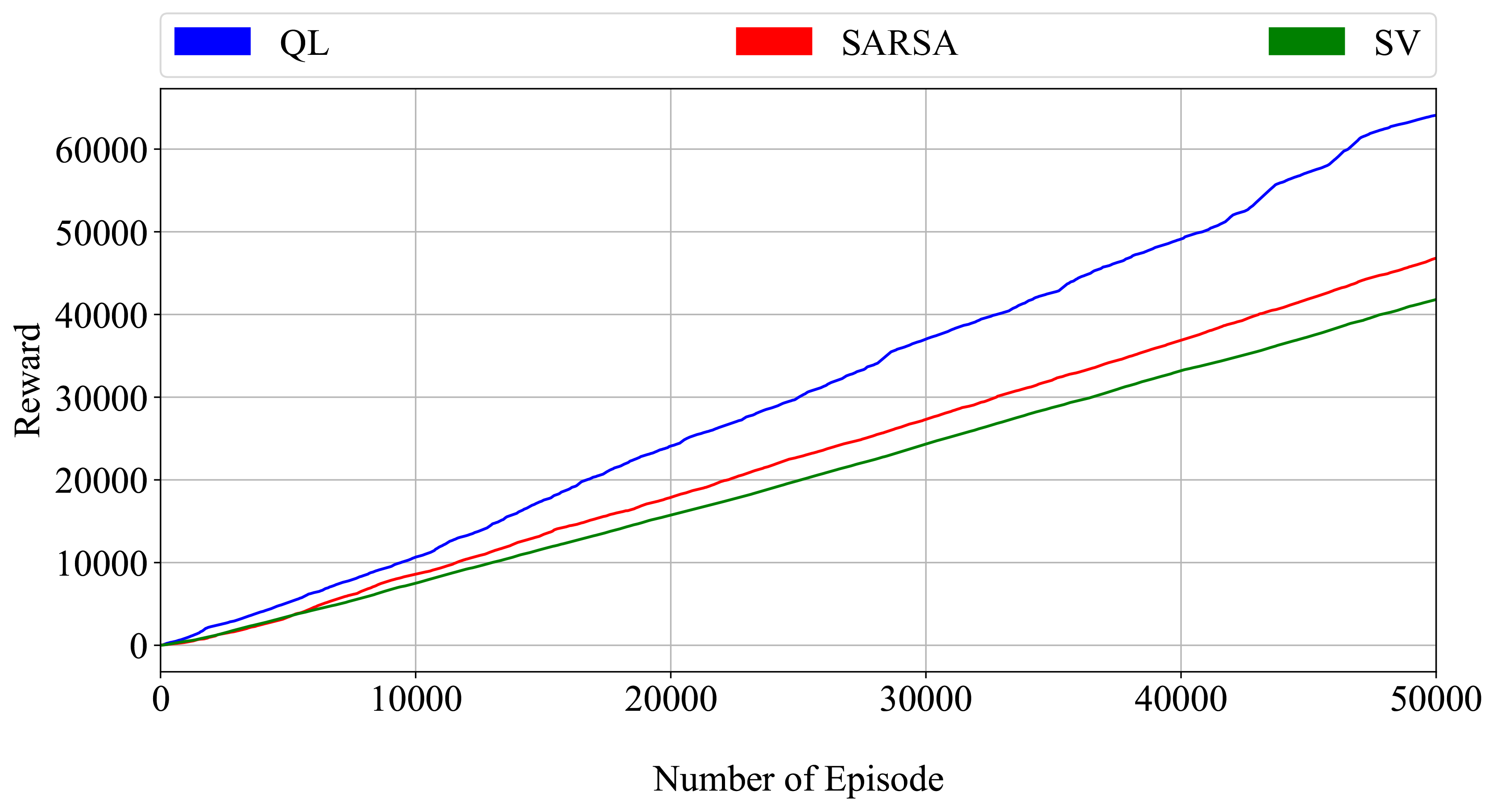}}\hfill
  \caption{Accumulated reward for 50 K episodes for Q-learning (QL), SARSA, and State-value (SV).} \label{fig_gen}
  \label{fig:Reward}
\end{figure}

All three approaches to the TD learning were implemented to Tic-tac-toe game scenario: QL, SARSA, and SV function evaluation. The algorithm was trained for 50 K episodes with parameters presented in Table \ref{tab:rl}. The reward values are accumulated by each approach for the training period and presented in Fig. \ref{fig:Reward}. 

\begin{table}[]
\caption{\footnotesize\scshape Parameters for the RL Learning}
\begin{center}
  \begin{tabular}{|p{2cm}|c|c|c|}
  \hline
  \renewcommand{\arraystretch}{1}
  \multirow{4}{*}{   Reward   } & \diagbox{Outcome}{Turn} & \multicolumn{1}{l|}{First turn} & \multicolumn{1}{l|}{Second turn} \\ \cline{2-4} 
   & Win & \multicolumn{2}{c|}{1} \\ \cline{2-4} 
   & Lose & \multicolumn{2}{c|}{-1} \\ \cline{2-4} 
   & Draw & 0.1 & 0.5 \\ \hline\hline 
  \multicolumn{2}{|c|}{Learning rate} & \multicolumn{2}{c|}{0.2}\\ \cline{1-4} 
  \multicolumn{2}{|c|}{Discount rate} & \multicolumn{2}{c|}{0.9}\\ \cline{1-4} 
  \multicolumn{2}{|c|}{Epsilon-greedy policy} & \multicolumn{2}{c|}{0.3}\\ \cline{1-4} 
  \end{tabular}
  \label{tab:rl}
  \vspace{-1.5em}
\end{center}
\end{table}


The results revealed that the QL algorithm performed faster than SARSA and SV approaches: QL performed in 38.8 sec, SARSA in 39.4 sec, SV in 87.6 sec on average per 10 K episodes. Additionally, the QL algorithm demonstrated on average 28\% higher reward in every episode than the SV method and 33\% higher reward than SARSA, winning 32\% of the time while learning. Therefore the trained QL algorithm was implemented for further experiments.

\subsection{Game Theory Algorithm}


To estimate and compare the user experience of interaction with a swarm of drones guided by the developed RL algorithm during the Tic-tac-toe game, we have adjusted a Basic Algorithm strategy \cite{Karamchandani}, applying it for the human-drone interaction scenario. 


Since the proposed drone-based scenario of the Tic-tac-toe requires more preparation time and complex actions from the swarm, in this research, we hypothesized that the high complexity of the game and deterministic strategy would not meet the player's expectations. To provide a considerable challenge and excitement for the user, we propose an Improved Basic (IB) Algorithm (see Algorithm 1).


\begin{algorithm} [h]
\caption{Improved Basic Algorithm}
\While {game is not over}
 {
    human turn\;
    
    checking for winning condition
    
    \eIf{drones can win (row, column, diagonal)}
    {
    insert “X" in the third cell and end game\;
    }
      {
        \eIf{human can win (row, column, diagonal)}
        {
        playing defensive\;
        }
        {
        play to make 2 in a row, column, diagonal (50 \%)\; 
        or random choice (50 \%)\; 
        }
      }
    {
    drone moving command\;
    
    checking for winning condition\;
    }
    \If {drones move first}
    {
    making the first move (50 \%)\;
    or random choice (50 \%)\;
    }
    \If {winning condition}
    {
    end game\;
    }
  }

\end{algorithm}

The purpose of this research is to achieve an interactive and realistic game process with the swarm. To simulate the errors caused by the human factor, we proposed a 50\% chance of a random move in the human winning scenario, when the IB algorithm can randomly place a drone into the empty cell at the start of the game, even if this is not going along its winning strategy. Therefore, the random factor increases the variability of the matches and allows the player to win in both game modes, providing an overall positive game experience.

\section{Experimental Evaluation}
\subsection{Research Methodology}
\subsubsection*{Participants} 


 
We invited 20 participants aged 22 to 43 years (mean = 25.6, std = 4.7, 4 females) to complete the survey. Two of them have never interacted with drones before, 5 of them regularly deal with drones, and 13 of them have participated in drone-based scenarios several times. 

\subsubsection*{Procedure} 
  
At the beginning of the experiment, the procedure and game equipment were introduced to each participant. Rules of the Tic-tac-toe were described for 1 participant who has never played the game before. 
Game elements, i.e., Noughts for human-player and Crosses for the SwarmPlay system, were represented by cardboard plates with printed black circles and cross shapes of the drones, respectively. Players placed the playing elements on the horizontally arranged game board, 1 m by 1.2 m white board with grid lines. 
All participants were randomly separated into two groups by 10 and played 2 matches with the SwarmPlay. The first group played the game with RL-based algorithm and the second with IB algorithm. 

At the end of the game, all respondents were asked to evaluate the SwarmPlay game with a Questionnaire based on a 5-point Likert scale. Both algorithms were evaluated separately on 7 metrics proposed in previous research on augmented games \cite{Kulshreshth_2013} and human-robot interaction systems \cite{Moschetti}: excitement, engagement, latency, challenge, tiredness, stress factor, and Turing test. A “Turing test” metric was proposed as an additional parameter to evaluate user's perception of their opponent as a person and not as an artificial system.
The participants responded to the following post-game questions:


  \begin{itemize}
     \item Excitement: How did you enjoy playing the game? Would you play it again? (Never again - Definitely).
     \item Engagement: To what extent did the game hold your attention? (Couldn't concentrate - Enjoyed the game).
     \item Latency: How appropriate did you find the responding time of the drones? (Unbearable - Unnoticeable delay).
     \item Challenge: To what extent did you find the game challenging? (Too easy - Really challenging).
     \item Tiredness: Did you feel tired after playing one game? (Not tired - Exhausted).
     \item Stress factor: Did you feel confident playing in the same space with drones? (Relaxed - Really stressful).
     \item Turing test: Did you consider your opponent being a person? (Completely artificial - It feels like playing with a human).
   \end{itemize}

\subsection{Experimental Results}

We conducted a chi-square analysis, based on the frequency of answers in each category.
The results showed that the game parameters are all independent (min RL: $p$ = 0.12 $\textgreater$ 0.05, min IB: $p$ = 0.16 $\textgreater$ 0.05).
Additionally the chi-square test of independence 
revealed that the participants' experience with drones does not affect the evaluation of drone swarm perception criteria, such as tiredness (RL: $\tilde{\chi}^2$ = 1.88, $p$ = 0.76, IB: $\tilde{\chi}^2$ = 5.60, $p$ = 0.47), stress factor (RL: $\tilde{\chi}^2$ = 3.69, $p$ = 0.45, IB: $\tilde{\chi}^2$ = 2.38, $p$ = 0.88) and Turing test (RL: $\tilde{\chi}^2$ = 6.03, $p$ = 0.19, IB: $\tilde{\chi}^2$ = 14.0, $p$ = 0.12). 




  \begin{figure}[htbp]
  \centering 
  \includegraphics[width=1\linewidth]{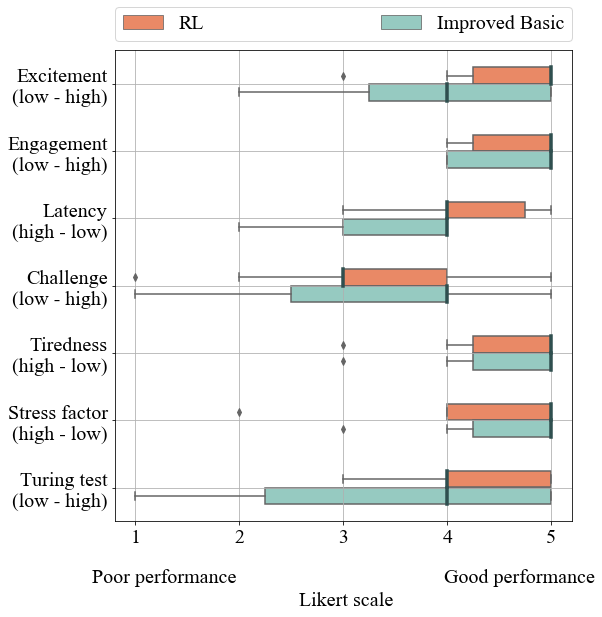}
    \caption{Questionary feedback on 5-point Likert scale for the human-drone game with RL-based algorithm and IB game theory algorithm.}
    \label{fig:statistics}
  \end{figure}


In summary, 70\% of the participants who played the IB Algorithm found the game exciting and expressed their interest in playing such games more (Excitement $\geq$ 4.0), almost 90\% of them did not feel any discomfort playing along with drones (mean score 4.6) and 60\% of users found the SwarmPlay response fast enough ($\geq$ 4.0) with game theory Algorithm compared to usual human-opponent move (mean score is of 3.6). 

The participants were fully engaged in the game based on the game theory algorithm (60\% put 5 score, 40\% - 4 score, mean is of 4.6), and about 90\% claimed they did not get tired playing with drones ($\geq$ 4, mean score 4.6). However, for the Turing test IB algorithm shows high variation 95\% CI (2.27, 4.6). 
Only 20\% of participants considered that playing with a robotic opponent was much distinguishable from the real person (Turing test evaluated as $\leq$ 3). Almost 70\% of respondents considered the game being challenging (challenge level evaluated as $\geq$ 4).

The results revealed the RL algorithm being more exciting for the participants, with 70\% of respondents put 5 score for RL algorithm, and only 30\% put 5 score for game theory algorithm (mean score IB: 3.9, RL: 4.6). The two-way \textbf{ANOVA} analysis results showed a statistically significant difference between the Excitement evaluation for RL-based algorithm and IB algorithm ($F$ = 4.05, $p$ = 0.047 $<$ 0.05).
The evaluation  of engagement level and tiredness for RL algorithm are close to IB (mean score Engagement RL: 4.7, IB: 4.6; mean score of Tiredness for RL: 4.6, for IB: 4.6). The results of the study is presented in Fig. \ref{fig:statistics}. 
However, for the Turing test RL algorithm showed less variation than IB (RL: 95\% CI(3.82, 4.78), IB: 95\% CI(2.32, 4.68)). Being evaluated more evenly the RL algorithm is proved to be closer to a real person, making intelligent moves, with some unpredictable for the opponent strategy. 
According to the received data, SwarmPlay won almost equal number of matches when drones started the game for 2 algorithms (40\% IB and 30\% RL), while only once the SwarmPlay won with IB and almost 30\% with RL-based algorithm when human-player moved first. SwarmPlay more often prevents human from winning the game with RL-based algorithm: 70\% Draw result and 30\% SwarmPlay won result with human player's first move. 
With applied game theory algorithm Draw outcomes occurred more frequently when SwarmPlay started (60\%) comparing to 30\% of draw outcomes with human-player start. For RL algorithm Draw result appeared equally often (70\%). 
The most interesting observation is that when human-players started, they won in 60\% games with IB and never with RL algorithm.



  


    \subsubsection*{Discussion} 


The results revealed that the first move is important for the game outcome (Table \ref{tab:start-win}).
 
\begin{table}
\centering
\caption{\footnotesize\scshape Relation between First Move and Game Results with RL-based and IB Algorithms}
\begin{tabular}{|p{2cm}|p{0.6cm}|p{0.6cm}|p{0.6cm}|p{0.6cm}|p{0.6cm}|p{0.6cm}|}
\hline
\multirow{2}{*}{\backslashbox[24mm]{Result}{Start}} & \multicolumn{2}{c|}{\begin{tabular}[c]{@{}c@{}}SwarmPlay\\First move~~\end{tabular}} & \multicolumn{2}{c|}{\begin{tabular}[c]{@{}c@{}}Human\\First move~~\end{tabular}} & \multicolumn{2}{c|}{\begin{tabular}[c]{@{}c@{}}Average time\\(sec)~~\end{tabular}}  \\ 
\cline{2-7}
& RL & IB & RL & IB & RL & IB\\
\hline
SwarmPlay won& 3 & 4& 3  & 1                                                                           & 60.3 & 65.6\\ 
\hline
Draw~~& 7&6& 7  & 3& 76.5 & 71.0                                                                         \\ 
\hline
Human won~~& 0    & 0& 0  & 6 & 0    & 29.5\\ 

\hline\hline
Time (sec):  & 70.1 & 74.0 & 73.1 &  68.4 & 71.9 & 70.4
\\

\hline
\end{tabular}
\label{tab:start-win}
\vspace{-1em}
\end{table}


 At the same time, with the applied IB algorithm, the first player won 46\% cases, whilst with RL-based algorithm won only 18.2\%. The RL-based algorithm was proved to be more successful in cases when the SwarmPlay takes the second turn. According to the received data, SwarmPlay won 35\% of all matches when drones started the game, with only 7.7\% of them won against game theory algorithm and 27.3\% against RL algorithm. SwarmPlay more often prevents human players from winning the game with RL algorithm: 72.2\% of the Draw result and 27.3\% of the SwarmPlay winning result with human player's first move. 

\begin{figure}[htbp]
  \subfloat[ \label{fig:1a}]{\includegraphics[width=0.24\textwidth]{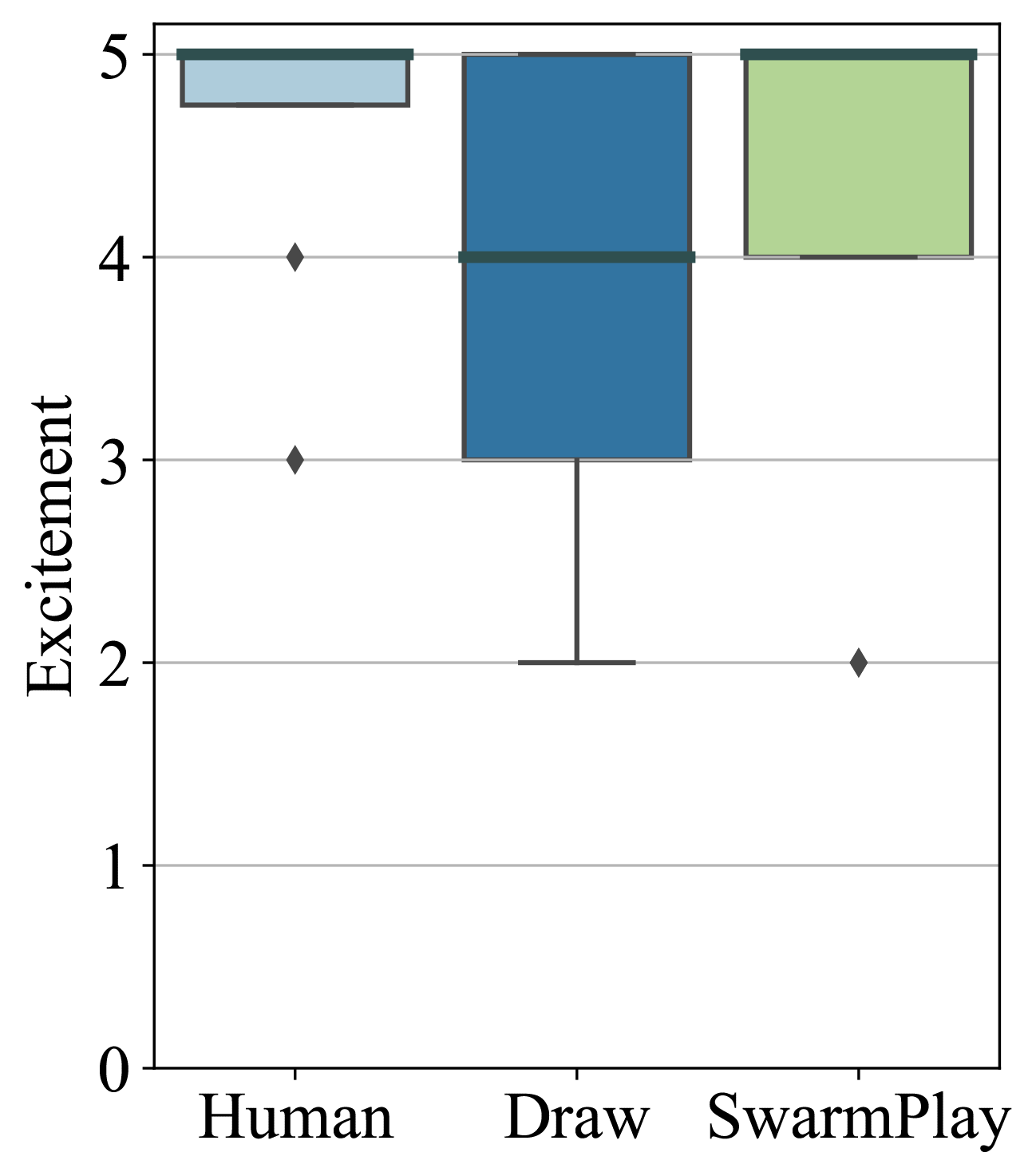}}\hfill
  \subfloat[ \label{fig:1b}] {\includegraphics[width=0.24\textwidth]{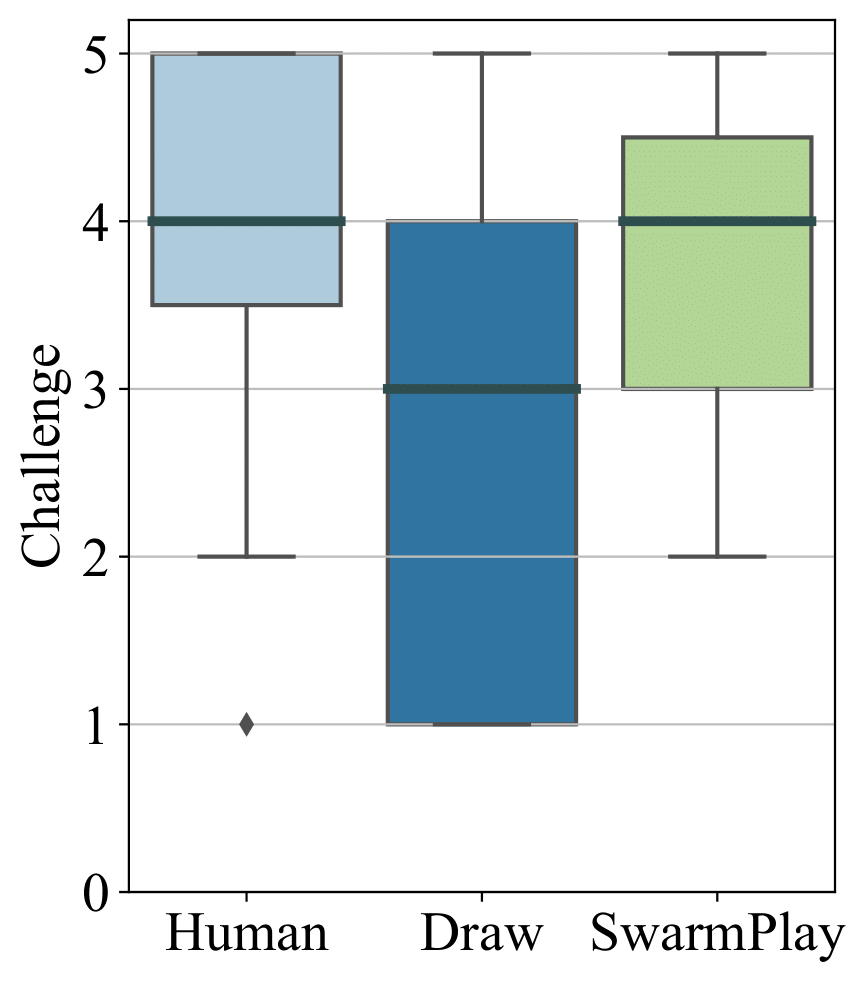}}\hfill
  \subfloat[ \label{fig:1c}]{\includegraphics[width=0.24\textwidth]{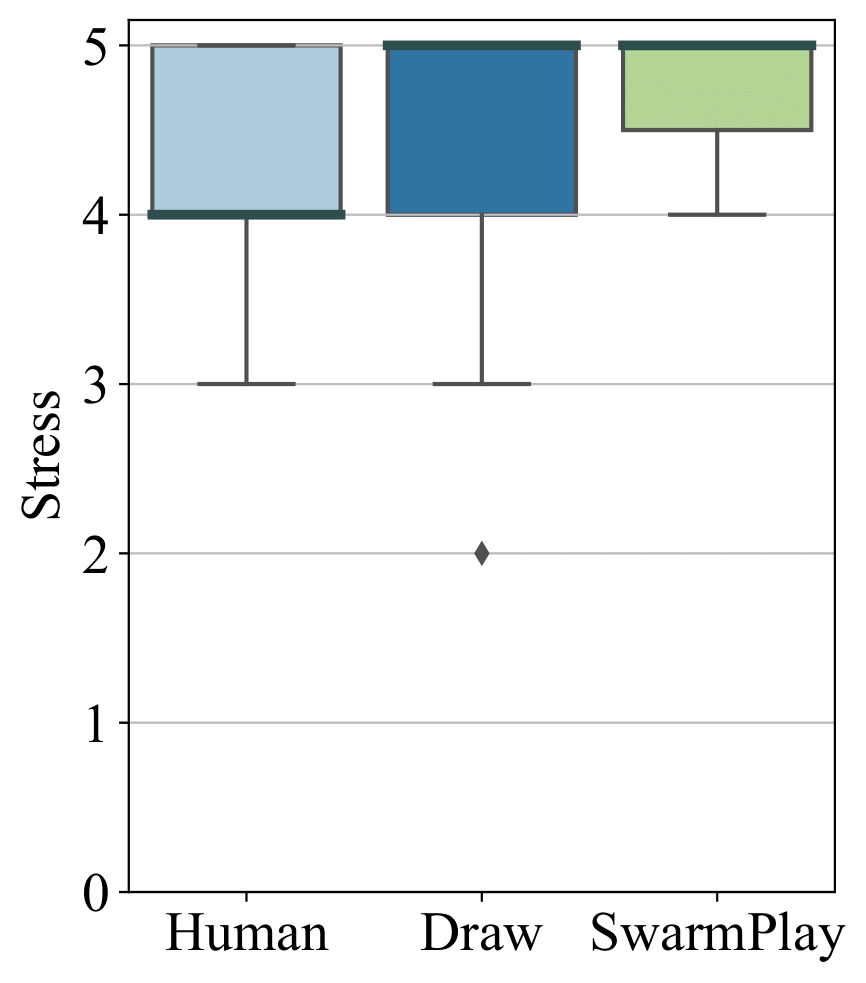}}\hfill
  \subfloat[ \label{fig:1d}]{\includegraphics[width=0.24\textwidth]{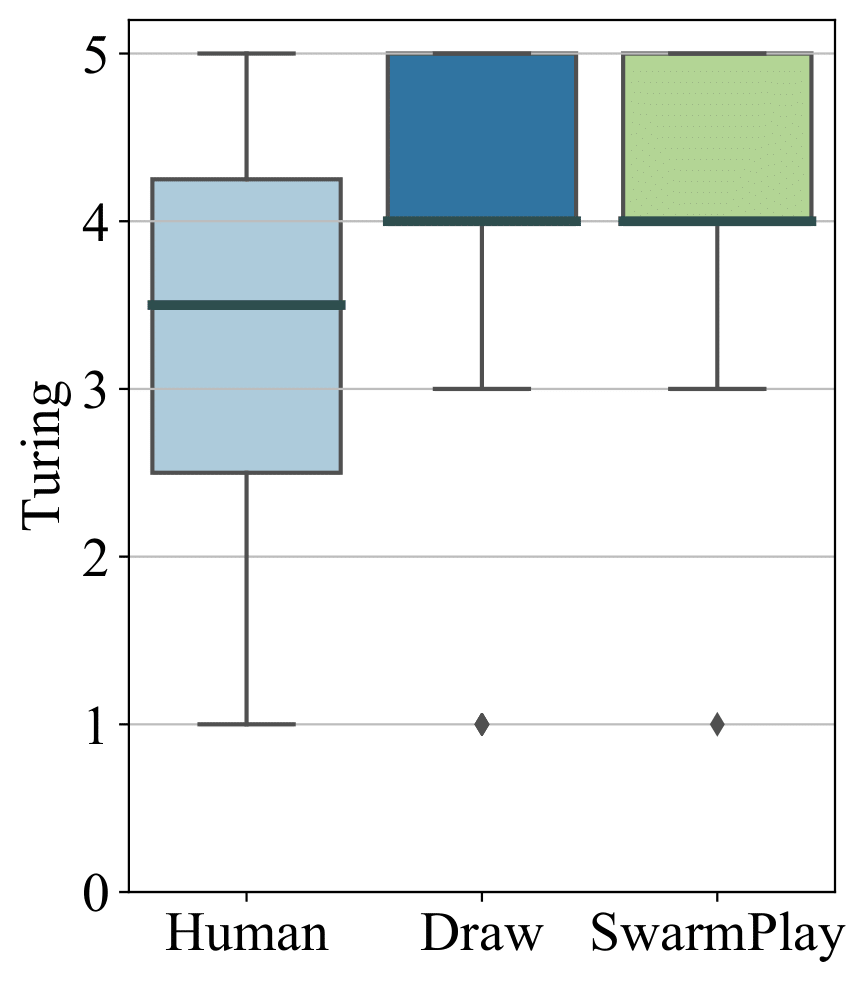}}
  
  \caption{Correlation between: (a) Excitement. (b) Challenge. (c) Stress. (d) Turing test on 5-point Likert scale and game outcome: human-player won, draw, or SwarmPlay won.} \label{fig:discussion}
\end{figure}

Surprisingly, we found a correlation between the game outcome (win-draw-lost) and game evaluation (Fig. \ref{fig:discussion}). 
Firstly, the human losses affect the user's excitement negatively (in average the excitement score is 12\% lower with Draw result and 10\% lower when SwarmPlay won than when a human player won). 
Nevertheless, users were much more focused on a game and not on system operation when SwarmPlay was leading the match. 
Secondly, it was discovered the more sophisticated strategy SwarmPlay performed and the more points it had the more human-like the behavior of the SwarmPlay participants mentioned. Meanwhile, results revealed that the participants evaluated the game challenge 34\% more when he or SwarmPlay won. 
Moreover, the participants felt more confident playing with drones when the game resulted in SwarmPlay win or Draw result (stress factor in average is 10\% less).

In the free comment space, four participants suggested implementing the external signal systems, e.g., sound alarms, either to indicate the start and the end of the game or to warn the player about swarm intentions. 



\section{Conclusions and Future Work}

We have developed the SwarmPlay, a system in which a human plays a Tic-tac-toe game against the swarm of drones. Our experimental results showed that 80\% of the participants found the game exciting and expressed their interest in playing the game again, wherein the RL algorithm was more exciting for the participants: 70\% of respondents put 5 scores for RL algorithm, and only 30\% for IB algorithm (mean score IB: 3.9, RL: 4.6). Participants showed a high engagement with the proposed technology (engagement mean score of 4.7 out of 5.0 for RL algorithm and of 4.6 for IB algorithm). 
Therefore, SwarmPlay can potentially improve our way of interaction with game pieces. Machines can not only learn from a human's winning strategy but also can teach humans how to achieve such a strategy throughout the interaction with an intelligent swarm.

The proposed system might be helpful in various scenarios, e.g., teaching the swarm of secure communication, where human shows the swarm places which can harm communication signals. Swarm then will establish the formations and communication nodes to achieve safe data transferring. Another approach could be open games with drones. The human-robot interaction via CV cameras can be implemented in training scenarios, to teach rescue personnel how to perform operations in a cluttered environment. There are a variety of home robot assistants, e.g., Kury, Buddy, Aido, etc. All these robots have a wheeled platform, therefore, they can move around the entire apartment. However, they can’t fly and bring humans any payload that is placed in cluttered and high locations. Home drone assistants can be used for that purpose. Humans can send the drones the target they need to reach, and they can do it in the safest way possible.


The future work will be devoted to more advanced board games, and we plan to apply ML techniques to learn the level of the player and adjust the difficulty level of the game in real-time. 


\bibliographystyle{IEEEtran}

\end{document}